\documentclass[a4paper,10pt]{article}

\usepackage{makeidx}         
\usepackage{graphicx}        

\usepackage{amsmath}
\usepackage{amssymb}

\usepackage[numbers,sort&compress]{natbib}

\makeindex             

\begin{document}

\title{Experimental Studies of Artificial Spin Ice}


\author{Christopher~H.~Marrows \\
School of Physics \&\ Astronomy, University of Leeds,\\ Leeds LS2 9JT, United Kingdom\\
Email: c.h.marrows@leeds.ac.uk}
%
%
\maketitle


\begin{abstract}
Artificial spin ices were originally introduced as analogs of the pyrochlore spin ices, but have since become a much richer field . The original attraction of building nanotechnological analogs of the pyrochlores were threefold: to allow room temperature studies of geometrical frustration; to provide model statistical mechanical systems where all the relevant parameters in an experiment can be tuned by design; and to be able to examine the exact microstate of those systems using advanced magnetic microscopy methods. From this beginning the field has grown to encompass studies of the effects of quenched disorder, thermally activated dynamics, microwave frequency responses, magnetotransport properties, and the development of lattice geometries--with related emergent physics---that have no analog in naturally-occurring crystalline systems. The field also offers the prospect of contributing to novel magnetic logic devices, since the arrays of nanoislands that form artificial spin ices are similar in many respects to those that are used in the development of magnetic quantum cellular automata. In this chapter, I review the experimental aspects of this story, complementing the theoretical chapter by Gia-Wei Chern.
\end{abstract}

\section{Introduction}
\label{sec:intro}

Whilst ice is a very commonplace material, it is never mundane: as I look out of my office window as I type these opening words on a fine February morning it coats the trees in a fine and beautiful frost; it will add a certain zip to the gin and tonic I'll pour to celebrate typing the closing sentence of this chapter. This is not to say that it lacks scientific interest: the complexity of its phase diagram and the wide variety of environments in which it is found mean that it is a very active area of interdisciplinary research \cite{Bartels-Rausch2012}.

To condensed matter physicists, one of its most remarkable properties must be that its most common form (Ice I$_\mathrm{h}$) has a residual entropy \index{residual entropy} at low temperatures due to proton disorder, and so violates the third law of thermodynamics. As Linus Pauling noted in 1935 \cite{Pauling1935}, the possible configurations of protons in Ice I$_\mathrm{h}$ provide a clear example of how geometrical frustration \index{frustration} can lead to vast numbers of energetically similar states, resulting in a non-zero entropy for temperatures arbitrarily close to absolute zero. This was one of the first recognised examples of a much broader phenomenon: frustration \index{frustration} occurs in a wide variety of condensed matter systems, and arises when it is not possible to simultaneously satisfy all of the competing pairwise interactions present. This leads to a rich phenomenology, where huge numbers of possible degenerate microstates play important roles in all kinds of complex systems in the physical sciences and beyond \cite{SteinandNewman}. Examples include liquid crystals, magnetic domain patterns, stripe structures in high-temperature superconductors, protein-folding, and neural networks.

Conventionally, the study of physical systems is restricted to the investigation of the limited set of naturally occurring materials. The family of rare earth pyrochlore materials closely resemble water ice \index{water ice} in their crystal geometry, and equivalent geometrical frustration \index{frustration} effects are found in the interactions between the large spins on the rare earth sites: hence they are dubbed `spin ices' \cite{bramwell2001}. One can mimic their behaviour using nanotechnology, which allows the construction of model systems where the nature of the elements and their interactions can be varied at will to create artificial frustrated systems \cite{Nisoli2013,HandSreview,chms_editorial}. Magnetic analogues of spin ice built using these methods, termed `artificial spin ice' \index{artificial spin ice} \cite{Nisoli2013,Wang2006}, have been widely studied in recent years since they provide convenient models for frustration \index{frustration} phenomena. One of the most appealing aspects is that advanced microscopy techniques are able to interrogate the system and reveal the exact configuration of all its constituent elements: yielding direct knowledge of the microstate (and the way that it evolves with time) of this artificial statistical mechanical system. When tied to the ability to engineer every aspect of the system during its construction, this is an extremely powerful new approach to the study of statistical mechanics, where experiments used to be restricted to revealing only the macrostate \index{macrostate}. In this chapter, an overview of experimental explorations of artificial frustrated systems is profiled. These studies have led to new insights into ordering and other dynamical processes in frustrated and disordered systems, and offer the prospect of technological innovations in information storage and processing.

\section{The First Spin Artificial Ices}
\label{sec:firstexpts}

In the initial, ground-breaking experiments from Schiffer's group at the Pennsylvania State University \cite{Wang2006}, artificial spin ice \index{artificial spin ice}s were conceived as exactly what their name describes: an artificially created system that mimics some aspects of the frustration \index{frustration} physics observed in the pyrochlore \index{pyrochlore} spin ices \cite{bramwell2001}. In the pyrochlore \index{pyrochlore}s, rare earth ions are found at the points where the corner-sharing tetrahedra touch. These ions carry large spin magnetic moments and strong crystal fields cause them to have a uniaxial anisotropy whose easy axis lies along the line between the centres of the two adjacent tetrahedra. Thus each moment must point into one of the two tetrahedra and out of the other one and so can take up only two states. In this sense the moments are described as being Ising \index{Ising}-like. When we consider all four moments in a given tetrahedron, it becomes obvious that their interactions must be frustrated, since it is not possible for them all to simultaneously form in-out pairs. The best that can be done is to form a `two-in/two-out' arrangement, identical to the Bernal-Fowler ice rule \index{ice rule} for the frustrated arrangement of protons in water ice \index{water ice} \cite{Bernal1933}. In this way the statistical mechanics of the spin ice are identical to that of Ice I$_\mathrm{h}$, and these systems have exactly the same extensive degeneracy and the same Pauling value \cite{Pauling1935} of the residual entropy \index{residual entropy} \cite{1999Ramirez}.

Schiffer's group constructed a system that is a physical realisation of the `square ice' \index{square ice} vertex models studied theoretically by theorists such as Wu \cite{Wu1967} and Lieb \cite{Lieb1967} in the 1960s, which were, in turn, versions of a model introduced by Slater for the ferroelectric material KH$_2$PO$_4$ \cite{Slater1941}. Their approach was to represent the rare earth spins in a pyrochlore \index{pyrochlore} spin ice by the macrospins \index{macrospin}s of nanomagnets. This required engineering the shape and size of the nanomagnets so that they were small enough to be in the single domain regime such that there was a well-defined macrospin \index{macrospin} and exhibited a clear uniaxial anisotropy to yield the required Ising-like\index{Ising} behaviour. To achieve these aims they selected 80~nm~$\times$~220~nm islands patterned from a 25~nm thick Permalloy \index{Permalloy} film. Permalloy \index{Permalloy} (Ni$_{81}$Fe$_{19}$) was selected since its lack of magnetocrystalline anisotropy means that its micromagnetics are entirely controlled by the shape of nanoelements: the rectangular shape provides the Ising-like \index{Ising} easy axis through shape anisotropy. The size is such that they form single domains and therefore behave as almost ideal Stoner-Wohlfarth particles \cite{1948Stoner}.

\begin{figure}[t]
  \includegraphics[width=8cm]{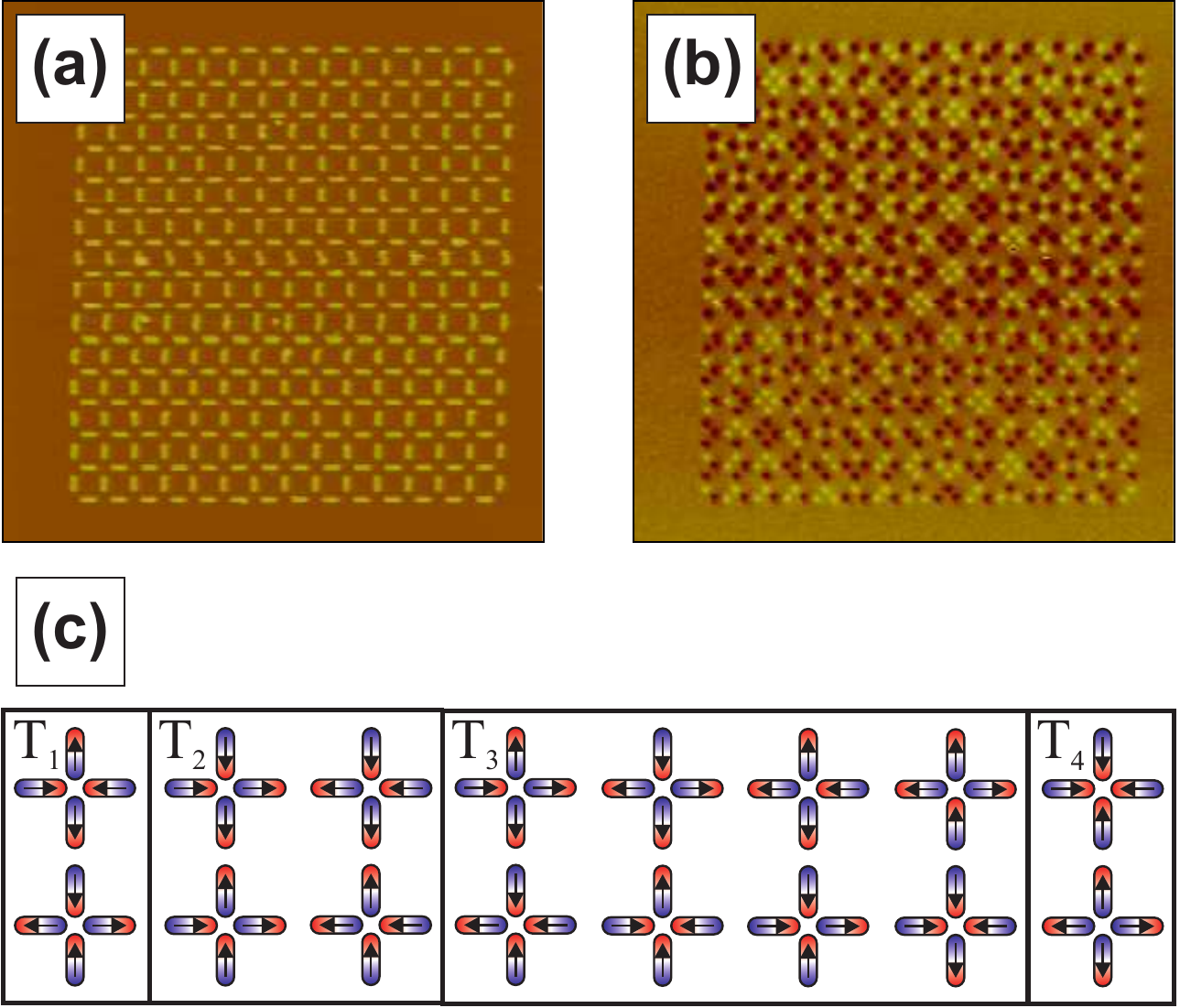}
  \caption{Square artificial spin ice. (a) Atomic force micrograph showing the arrangement of nanomagnets along the edges of a square tiling. (b) Magnetic force micrograph of the same area, showing the magnetic poles of each nanomagnet as bright or dark contrast. (c) The sixteen possible magnetic configurations of a vertex in a square spin ice, divided into four types by energy. \label{square_ice}}
\end{figure}

These elements are then arranged along the edges of a square tiling, such that they meet at the vertices of the tiling in groups of four, as shown in Fig. \ref{square_ice}. The mutual magnetostatic \index{magnetostatic} interactions between the four macrospins \index{macrospin} then mimic those between the four rare earth spins meeting in each tetrahedron of a pyrochlore \index{pyrochlore} spin ice. There are $2^4 = 16$ possible magnetic configurations of any given vertex, as shown in Fig. \ref{square_ice}(c), of which six obey the two-in/two-out ice rule \index{ice rule}. Whilst in the 3-dimensional pyrochlore \index{pyrochlore} system all six are energetically degenerate, here the reduction in dimensions means that not all the pairwise interactions are equivalent, and so these six can be further subdivided into a group of two with the lowest energy, and a further four with a slightly higher energy. (The interactions between neighbouring nanomagnets are stronger than those between nanomagnets that face each other across the vertex). In the terminology introduced by Wang \textit{et al.}, and subsequently adopted by the community, these two groups are referred to as `Type 1' and `Type 2', respectively. There is no net magnetic pole at the centre of these two types of vertices. The eight `Type 3' configurations are those where the ice rule \index{ice rule} is violated by flipping the moment of a single element, these have a `three-in/one-out' or `three-out/one-in' arrangement, and carry a net magnetic charge of $\pm 2q$, where $q$ is the magnetic charge of a single pole at the tip of one element. The two highest energy configurations, `all-in' or `all-out', comprise `Type 4', and carry a magnetic charge of $\pm 4q$. The Type 2 and Type 3 vertices also possess a net magnetic dipole moment. The fact that only two vertex types are selected for the lowest energy state means that this system possesses a twofold degenerate ground state \index{ground state} consisting of a chessboard tiling of the Type 1 vertices \cite{Morgan2011a}, as opposed to the macroscopic degeneracy of the pyrochlore \index{pyrochlore} system.

In the ideal case, studied theoretically as the six-vertex model \cite{Wu1967,Lieb1967}, only Type 1 and Type 2 vertices are found, and the ice rule \index{ice rule} is obeyed everywhere. On the other hand, in a non-interacting system, all sixteen vertex types will be found with equal probability, since they will be randomly populated. The experimental results of Wang \textit{et al.} fell between these two extremes \cite{Wang2006}. Whilst all sixteen vertex types were populated, the system showed a preference for those of Type 1 and Type 2, with an accompanying deficit in the populations of Type 3 and Type 4, as compared to the random case. This preference grew stronger as the lattice constant of the square ice \index{square ice} was reduced, packing the nanomagnets more closely together, and allowing the stronger magnetostatic \index{magnetostatic} interactions to better enforce the ice rules \index{ice rule}.

\begin{figure}[t]
  \includegraphics[width=8cm]{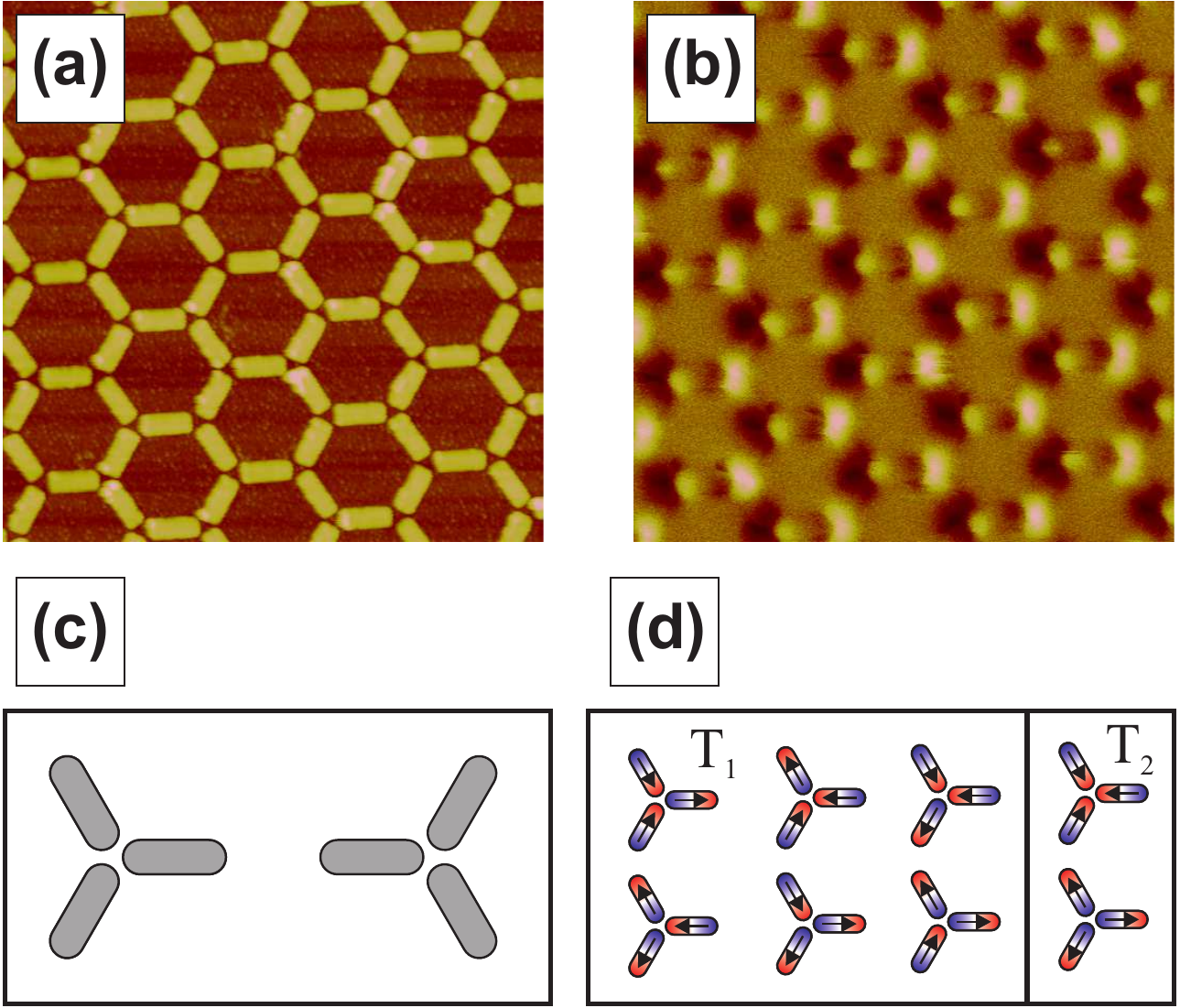}
  \caption{Kagome artificial spin ice. (a) Atomic force micrograph showing the arrangement of nanomagnets along the edges of a honeycomb \index{honeycomb} tiling. (b) Magnetic force micrograph of the same area, showing the magnetic poles of each nanomagnet as bright or dark contrast. (c) The two inequivalent structural vertex types of a kagome spin ice. (d) The eight possible magnetic configurations of a vertex in a kagome spin ice. \label{kagome_ice}}
\end{figure}

The other ice-like geometry studied at the outset of the field is the so-called kagome or honeycomb \index{honeycomb} ice, which had been analysed theoretically by Wills \textit {et al.} \cite{Wills2002}. Here the macrospins \index{macrospin} are arrayed along the edges of a hexagonal tiling (see Fig. \ref{kagome_ice}), visually resembling a honeycomb \index{honeycomb}. The lattice points at the centre of each element lie on a kagome lattice. This system is analogous to the pyrochlore \index{pyrochlore} spin ice under a large field applied along a $\langle 111 \rangle$ direction \cite{Matsuhira2002}. The fact that three elements meet at each vertex means that all the interactions at that vertex are of equal strength. Artificial experimental realisations of this geometry were studied in a connected form by Tanaka \textit{et al.} at Keio University in Japan \cite{2005Tanaka} and as small groups of discrete islands by Mengotti \textit{et al.} at the Paul Scherrer Institute in Switzerland \cite{Mengotti2008}. This lattice has two inequivalent structural vertex types, depicted in Fig. \ref{kagome_ice}(c). Each of these can take up $2^3 = 8$ possible magnetic configurations: Type 1 obeys a two-in/one out or one-in/two out pseudo-ice rule \index{ice rule} and carries a net charge of $\pm q$; Type 2 has an all-in or all-out configuration and carries a charge of $\pm 3q$. These are shown in Fig. \ref{kagome_ice}(d). There are no uncharged vertices in this structure.

\section{Experimental Methods}

The methods used in experimental studies of artificial spin ice \index{artificial spin ice} fall into two main categories: those used to fabricate the arrays of nanoscale magnets that form the artificial frustrated systems, and those used to study the magnetic states they take up, which primarily various forms of advanced magnetic microscopy.

\subsection{Fabrication Methods}
\label{sec:fabrication}

The fabrication methods for artificial spin ice \index{artificial spin ice}s are the conventional nanofabrication routes that are now well-established in magnetic nanotechnology \cite{Martin2003}. The critical lateral dimensions of the elements that form artificial spin ices are in the 10--100~nm range, and this means that electron beam lithography is the method of choice for writing the patterns that form the frustrated arrays. In this technique a pattern is written into an electron sensitive resist by raster-scanning a highly focussed electron beam across it, referred to as exposing the pattern. The resist is spin-coated onto the substrate and is typically a polymer that will undergo either scission (for a positive tone) or cross-linking (negative tone) upon electron beam exposure. This changes the solubility, and so the pattern may be developed by washing with an appropriate solvent. The pattern may then be transferred by depositing or etching material on the substrate surface. The natural variations in the exact shape and size of the islands corresponds to the quenched disorder in the statistical mechanical system that is being engineered \cite{Daunheimer2011}.

The use of a negative tone resist is most common, in a pattern transfer process referred to as `lift-off'. Here the resist is coated onto a bare substrate and once the pattern is developed the resist forms a stencil mask through which magnetic material may be deposited to form the desired nanostructure: in this case an array of islands. The use of a strong solvent then removes (lifts-off) all the resist and leaves behind only the material deposited through the holes onto the substrate. The most common resist used for liftoff is polymethylmethacrylate (PMMA). Many common magnetic materials can be deposited by thermal evaporation: Permalloy \index{Permalloy} falls into this group. this technique is easily compatible with lift-off, since it is a high vacuum process and there is thus ballistic transport of the deposited atom flux from the source to the substrate. This highly directional flux means that the sidewalls of the holes that have been opened in the resist are hardly coated at all and lifting-off the resist proceeds easily. Things are more complex when the other main deposition technology, sputtering, is used. Since sputtering involves the use of a working gas at a few mTorr pressure, the mean free path of the deposited flux is short and there is significant step coverage, meaning that clean liftoff is difficult. Two possibilities now present themselves. First, a bilayer resist that provides a large undercut can be used, meaning that the sputtered film cannot easily coat the sidewalls. On the other hand, patterns can be written and developed in resist spun on top of a sputtered film to provide an etch mask, with the pattern subsequently transferred by wet (chemical) or dry (ion mill or reactive ion) etching. The fact that the magnetic islands forming a spin ice array grow as discrete elements during liftoff has consequences for their magnetic configuration, since the macrospin \index{macrospin} system is able to thermalise during the early stages of growth \cite{Morgan2011a,Nisoli2012}. For transmission microscopy experiments (e.g. Lorentz imaging or X-ray transmission microscopy) it is necessary to carry the sample fabrication on an electron transparent substrate, typically a Si$_3$N$_4$ membrane.

\subsection{Measurement and Imaging Methods}
\label{sec:measure}

The experimental methods used to study artificial spin ice \index{artificial spin ice}s can be divided into two categories. The first are ``macroscopic'' probes which measure a quantity that is averaged over the entire sample: this can be thought of as measuring, in some sense, the statistical mechanical macrostate \index{macrostate}. On the other hand, one of the main advantages of the artificial spin ice \index{artificial spin ice} approach is that the lengthscales make the system amenable to study using the various advanced magnetic microscopy methods that have been developed over the past few decades. These allow the statistical mechanical artificial spin ice \index{artificial spin ice} to be interrogated, and its response to external stimuli probed.

Most of the macroscopic probes used to study magnetic materials have been applied, in some way, to the study of artificial spin ice \index{artificial spin ice}s. Whilst it is just about possible to write a pattern that is large enough for measurement using conventional magnetometry (e.g. by SQUID magnetometer), this is challenging, since it relies on a very long write time. A more convenient way to measure the magnetic hysteresis loop of a an object that is microns or tens of microns across--the typical size of an artificial spin ice \index{artificial spin ice} array--is to use the magneto-optical Kerr effect (MOKE) \cite{Kohli2011}, more elaborate forms of which include Bragg-MOKE in a scattering geometry \cite{2008Westphalen}, and a temperature dependent form used by Kapaklis \textit{et al.} \cite{Kapaklis2012}. If connected arrays are formed then they become able to conduct an electrical current and magnetotransport \index{magnetotransport} measurements become possible \cite{Tanaka2006,Le2015}, which are also able to probe switching fields and hence interactions within the array. The fact that the artificial spin ices \index{artificial spin ice} are periodic arrays means that soft X-ray resonant magnetic scattering (XRMS) can be used to study them, since they will give rise to well-defined diffraction peaks \cite{Morgan2012,Perron2013}. The generation of such soft X-rays requires a synchrotron. This method, as is usual with scattering techniques, gives spatial resolution in reciprocal space, and  allows for the sampling of a larger area of the sample than the typical field-of-view of a microscope. Last, it has been predicted that ferromagnetic \index{ferromagnet} resonance (FMR) measurements \cite{Sklenar2013} can also be used to study ordering in artificial spin ices \index{artificial spin ice} \cite{Gliga2013}.

\begin{figure}[t]
  \includegraphics[width=8cm]{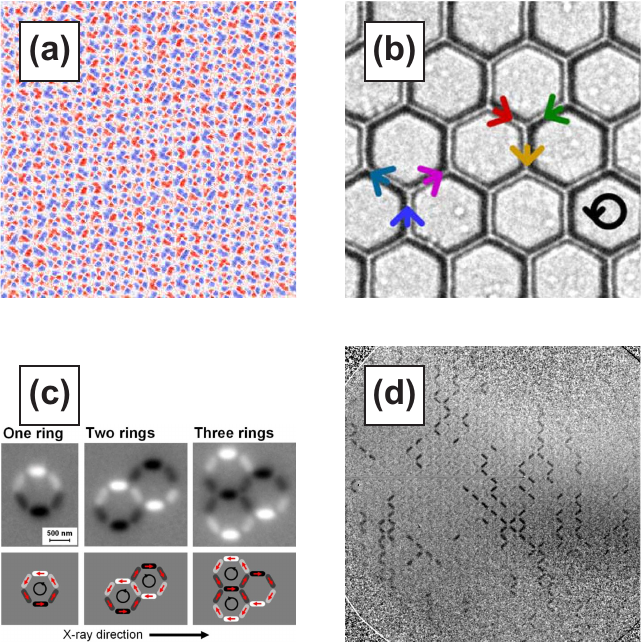}
  \caption{Magnetic micrographs of artificial spin ices, obtained by (a) magnetic force microscopy, where blue/white contrast corresponds to opposite vertical force gradients, revealing the charges in this square ice array, (b) Lorentz microscopy, here in Fresnel mode, where dark/bright contrast on the sides of the bars in this connected kagome network reveals the magnetisation direction (after Qi \textit{et al.} \cite{2008Qi}), (c) photoemission electron microscopy, where the contrast depends on XMCD and therefore is given by the relative alignment of the magnetisation and the X-ray beam direction, here used to reveal the magnetic states of small hexagonal rings that are kagome ice prototypes (after Mengotti \textit{et al.} \cite{Mengotti2008}) and (d) Full field X-ray transmission microscopy, which also uses XMCD as the effect to yield contrast, showing Dirac strings of reversed elements. \label{micrographs}}
\end{figure}

Of course, imaging the exact magnetic configurations of artificial spin ices \index{artificial spin ice} leads to deeper  insights into their behaviour. Since the individual elements are of a size that is smaller then the wavelength of visible light, they cannot be resolved in a optical microscope and so techniques such as Kerr microscopy cannot be used. The simplest method by which the magnetism in objects that are a few tens or hundreds of nm in size can be observed is magnetic force microscopy MFM (\index{MFM}). This is a variant of the atomic force microscope method, where the tip is magnetised and so responds to magnetic, as well as van der Waals forces. These are generated by field gradients, and so this method is well-suited to revealing the poles at the ends of the nanomagnets, as shown in Fig. \ref{micrographs}(a). From these, the direction of each macrospin \index{macrospin} is easily determined. This method was the one used in the first experiments at Penn State \cite{Wang2006} and remains popular due to the wide availability of such microscopes \cite{Ladak2010,schumann2010,Lammert2010,Nisoli2010,Morgan2011a,Morgan2011b,Porro2012,Zhang2013,Gilbert2014}. The typical spatial resolution is $\sim 50$~nm, although more advanced instruments can do better than this. A potential drawback is that the magnetic tip can influence the magnetic state of the sample as it is scanned across it if that state is particularly labile. Measurements can, in principle, be carried out under an applied field, although most MFM \index{MFM} instruments are usually AFMs retrofitted with a magnetic tip, and so usually only specialised instruments are equipped with magnets.

The other imaging technique used at the outset of the field was Lorentz microscopy \cite{Tanaka2006}. This is a form of transmission electron microscopy, and so has very high spatial resolution--typically better than 10~nm, but can be up to an order of magnitude less in modern aberration corrected microscopes--but requires the artificial spin ice \index{artificial spin ice} array to be on a membrane that is thin enough to be electron transparent: typically 50-100~nm of Si$_3$N$_4$. This means, of course, that such samples are very fragile. Contrast arises due to the deflection of the electrons by Lorentz forces as they pass through the sample, which can be accessed if the image is deliberately slightly defocussed. Imaging can be carried out under an applied field. An example of a Lorentz micrograph is shown in Fig. \ref{micrographs}(b). It has therefore proved popular among those groups with suitable transmission electron microscopes \cite{2008Qi,Phatak2012,Daunheimer2011,Pollard2012,Shen2012}. Since this method depends on the magnetic flux \textbf{B} that the electron beam passes through for contrast, it is possible to image the flux-lines between the elements that give rise to the frustrated couplings. This has been accomplished using the related electron holography technique \cite{Phatak2011}.

The other main imaging technique used in the field of artificial spin ices \index{artificial spin ice} is synchrotron-based: photoemission electron microscopy (PEEM \index{PEEM}) using X-ray magnetic circular dichroism (XMCD) as the contrast-generating mechanism. Here the sample is illuminated with soft X-rays tuned to an absorption edge of a ferromagnetic \index{ferromagnet} element: for the most commonly used material, Permalloy \index{Permalloy}, the $L_3$ edge of Fe is used. The XMCD effect is connected to the XRMS through a Kramers-Kronig relation, and means that there is differential absorption of the X-ray photons depending on whether the photon helicity is parallel or antiparallel to the magnetisation. This differential absorption in differently oriented domains leads to different photoelectron yields, giving rise to a magnetic contrast mechanism in an image formed when those photoelectrons are collected in an electron column. The fact that these photelectrons are easily Lorentz deflected means that employing this technique precludes the application of magnetic fields to the sample during imaging. The typical spatial resolution is better than $\sim 100$~nm. An example of an XMCD-PEEM \index{PEEM} micrograph is shown in Fig. \ref{micrographs}(c). The use of this method as a means of studying artificial spin ices \index{artificial spin ice} was pioneered by Heyderman's group at the Paul Scherrer Institute \cite{Mengotti2008}, and it has subsequently been adopted by others \cite{Mengotti2011,2011Rougemaille,Rougemaille2013,Pushp2013,Montaigne2014}. The extreme surface sensitivity of the technique, arising from the short (2-3~nm) escape depth for the photoelectrons, means that it can be used to study extremely thin samples that cannot be imaged by the other methods described here. This means that the volume of the nanoislands is very small, allowing them to be influenced by thermal fluctuations \index{fluctuation} \cite{Farhan2013a,Farhan2013b,Kapaklis2014}.

Transmission X-ray microscopy (TXM) combines features of the Lorentz TEM and PEEM \index{PEEM} methods. Like the former, it is a transmission method and requires samples in the same type of thin membranes. Like the latter, it makes use of XMCD as a contrast generating mechanism. However, since it is a photon-in/photon-out method, it can be performed with the sample under an applied field, and so can be used to track the response of artificial spin ice artificial spin ice \index{artificial spin ice} arrays as they are taken around their hysteresis loops. An example of an XTM micrograph is shown in Fig. \ref{micrographs}(d). It has been most commonly used in the scanning XTM (STXM) implementation to study artificial spin ice \index{artificial spin ice} samples, where a focussed soft X-ray is scanned over the sample  \cite{Ladak2012,Zeissler2013,Walton2015}.

\section{Monopoles and Magnetricity}
\label{sec:monopole}

A paradigm-defining result in spin ice physics was the development by Castelnovo \textit{et al.} of the theoretical description of excitations \index{excitation} in pyrochlore \index{pyrochlore} spin ices in terms of deconfined magnetic monopoles \index{monopole} \cite{Castelnovo2008}. The central concept is that whilst a vertex in the tetrahedral pyrochlore \index{pyrochlore} network possesses no net magnetic charge when the two-in/two-out ice rule \index{ice rule} is obeyed, a violation of that ice rule \index{ice rule}, caused by a spin-flip, leads to the creation of net positive and negative magnetic charges in the two tetrahedra connected by that spin. This can be thought of as corresponding to a monopole-antimonopole \index{monopole} pair creation event. Flipping adjacent spin, then another, and another, gradually separates the monopole \index{monopole} and antimonopole \index{monopole}, with the chain of flipped spins being analogous to the flux-tube known as a Dirac string \index{Dirac string} in the theory of fundamental cosmic monopoles \index{monopole} \cite{Dirac1931}. An important feature of the pyrochlore \index{pyrochlore} geometry is that once this has been done, whilst the sites of the oppositely charged monopoles \index{monopole} can be easily seen, it is not possible to tell which chain of spins has been flipped to separate them. As a result there is no string tension and the energy of the monopoles \index{monopole} is determined only by their separation through an equivalent of the Coulomb law. This picture was soon experimentally confirmed using neutron \cite{Fennell2009,2009Morris,2009Kadowaki} and muon \cite{2009Bramwell} experiments on different pyrochlore \index{pyrochlore} systems. Building upon the concept of mobile magnetic charge carriers, the idea of `magnetricity' \index{magnetricity} has been introduced. This is the creation and flow of mobile magnetic charges that respond to externally applied magnetic fields \cite{Giblin2011}.

Translating this picture into the language of square spin ice, as introduced in \S \ref{sec:firstexpts}, we can see that whilst Type 1 and Type 2 vertices are charge neutral (obeying the ice rules \index{ice rule}), the monopoles \index{monopole} in the Castelnovo \textit{et al.} picture correspond to the charged Type 3 vertices. A type 4 vertex corresponds to the superposition of two monopoles \index{monopole} of the same charge. Since the Dirac strings \index{Dirac string} must carry flux, they must be composed of the dipole-bearing Type 2 vertices. The fact that there is a distinction to be drawn in a square spin ice between the Type 1 and 2 vertices means that in this case a unique Dirac string \index{Dirac string} can be identified \cite{2009Moller}. Under these circumstances the monopoles \index{monopole} are not deconfined. For this reason, strictly speaking, the monopoles \index{monopole} present in square artificial spin ice \index{artificial spin ice} systems \cite{Silva2013} are of the Nambu type \cite{Nambu1974}. In a kagome spin ice, all vertices must bear charges, since an odd number of magnetic dipoles meet at each vertex. The same monopole/Dirac string \index{monopole} \index{Dirac string} picture applies in this case \cite{Ladak2010}, but the monopoles \index{monopole} must be defined as charge differences with respect to a well-defined state \cite{Mengotti2011}.

\begin{figure}[t]
  \includegraphics[width=9.2cm]{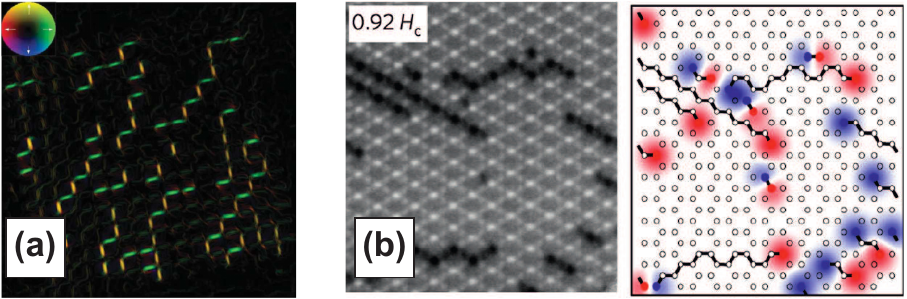}
  \caption{Magnetic monopoles and Dirac strings in (a) square artificial spin ice, visualised using Lorentz microscopy (after Pollard \textit{et al.} \cite{Pollard2012}) and (b) kagome artificial spin ice, visualised using PEEM, with an accompanying schematic (after Mengotti \textit{et al.} \cite{Mengotti2011}). \label{monopoles}}
\end{figure}

The observation of monopoles \index{monopole} and their field-driven dynamics in artificial ices was in fact first accomplished in the kagome systems \cite{Ladak2010,Mengotti2011}. The premise of the experiment is simple: the magnetic array is first saturated in one direction and then a reverse field applied to take it along one branch of its hysteresis loop. (In fact, in these two experiments, due to the constraint of only being able to image at zero field, were in fact measurements of the first-order reversal curve, although the thermal bistability of the systems means that this is equivalent to a major hysteresis loop in this case.) Mengotti \textit{et al.} performed an analysis of the size distribution of the Dirac strings \index{Dirac string} that they observed in their experiments on a kagome ice \index{kagome ice} formed from discrete Permalloy \index{Permalloy} islands \cite{Mengotti2011}. The propagation of these Dirac strings \index{Dirac string} may be regarded as avalanche events. It is usually the case, in physics and beyond, that the distribution of avalanches sizes follows a power law, often referred to as a Gutenberg-Richter law \index{Gutenberg-Richter law} (terminology that first arose in seismology), and so is scale free. In contrast to this, an exponential law was found, implying a characteristic scale. This was attributed to the fact the Dirac strings \index{Dirac string} are 1-dimensional objects in a 2-dimensional system, viz. the frustrated interactions enforce a lowering of the dimensionality \index{dimensionality} of the events. This general type of reversal (nucleation of monopole-antimonopole \index{monopole} pairs, followed by Dirac string \index{Dirac string} cascades) has also been shown to take place in artificial square spin ices \cite{Pollard2012,Morgan2011b}. Micrographs showing Dirac strings \index{Dirac string} in both geometries are displayed in Fig. \ref{monopoles}.

Whereas the studies discussed above are performed for monopoles \index{monopole} constructed from the uncompensated magnetic poles at the ends of finite-sized discrete elements (which do show monopole-like \index{monopole} field distributions \cite{Phatak2011}), similar concepts can be used to interpret the behaviour of connected spin ice arrays, which are networks constructed from magnetic wires. These vertices emerge at the intersections of the magnetic wires usually studied in the form of a honeycomb \index{honeycomb} lattice equivalent to a kagome ice \index{kagome ice} \cite{Ladak2010,Ladak2012}. In these systems, monopoles \index{monopole} are created by the emission and interaction of magnetic domain walls \index{domain wall} at the wire intersections. This is expected to be governed by the chirality and topological properties of the domain walls \cite{Pushp2013}, although this has been recently questioned \cite{Walton2015}, since quenched disorder in the system may affect this ideal behaviour in samples where the lithographic defects are not under complete control.

\section{Array Topology and Geometry}
\label{sec:geometry}

It is a truism that in geometrically frustrated systems, control of both the static states and their dynamics can be achieved through a proper design of the topology and geometry of the network of interacting elements. So far in this chapter we have discussed only the two most common geometries, square and kagome, which mimic different aspects of the tetrahedral pyrochlore \index{pyrochlore} systems. Nevertheless, one of the main attractions of the artificial spin ices \index{artificial spin ice} approach is that the array geometry can be designed and engineered lithographically, and so the only limitation on possible designs is the imagination of the designer.

\begin{figure}[t]
  \includegraphics[width=10cm]{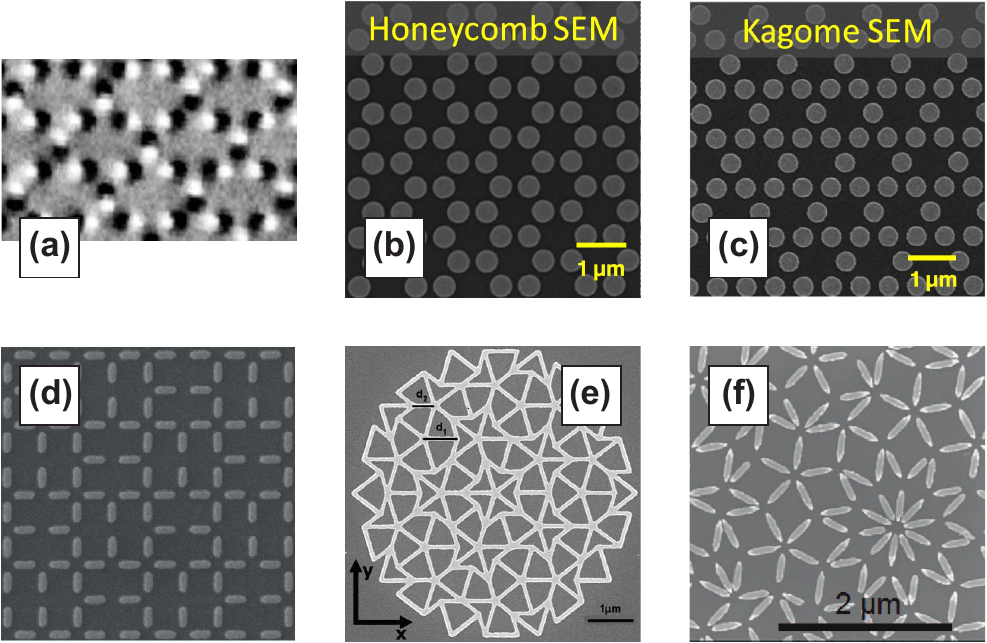}
  \caption{Novel geometries for artificial spin ices include (a) the brickwork lattice (MFM, after Li \textit{et al.} \cite{Li2010}), (b) hexagonal and (c) kagome networks of perpendicularly magnetised islands (SEM, both after Zhang \textit{et al.} \cite{Zhang2012}), (d) the shakti lattice (SEM, after Gilbert \textit{et al.} \cite{Gilbert2014}), and (e)  kite-and-dart (SEM, after Bhat \textit{et al.} \cite{Bhat2013}) and (f) rhomboid quasiperiodic Penrose tilings (SEM). \label{geometries}}
\end{figure}

A so-called `brickwork' \index{brickwork} lattice was developed by the Penn State group, depicted in Fig. \ref{geometries}(a) \cite{Li2010}. It is constructed by removing staggered rows of every other element from a square ice \index{square ice} system. This maintains the axes of the islands to be orthogonal to each other, just as for square ice \index{square ice}, but reduces the number of islands meeting at each vertex to three, making it topologically equivalent to the kagome lattice. However it retains the property of the square lattice that the interactions at each vertex are not all equivalent, since the angles between the nanoislands are not all the same. As a result the properties of the lattice, such as approach to the ground state \index{ground state} on ac demagnetisation \index{demagnetisation} and the correlations between the island alignments, are more similar to the square than kagome structure. The details of the geometry matter in artificial frustrated systems: they cannot be classified on topology alone.

Thin magnetic films are usually magnetised in-plane since they have a strong shape anisotropy, and it is usually this anisotropy that is used to provide the Ising-like \index{Ising} nature of the macrospins \index{macrospin} in artificial spin ices \index{artificial spin ice}. Nevertheless, other anisotropies can be used. Perpendicular anisotropies are common at the interfaces between magnetic and heavy metals due to the lowering of symmetry at a point where there are strong spin-orbit interactions. For a sufficiently thin ferromagnetic \index{ferromagnet} film, which in practice usually means $\lesssim 1$~nm, the reduction in the volume demagnetising energy is large enough that this interface anisotropy can dominate and the magnetic easy axis is out-of-the-plane. A prototypical materials system displaying this effect is a multilayer \index{multilayer} of Co/Pt. A nanoscale dot of such a multilayer \index{multilayer} will possess an Ising-like \index{Ising} macrospin \index{macrospin} that is directed out-of-the-plane and so the magnetostatic \index{magnetostatic} coupling to a neighbouring dot will be antiferromagnetic \index{antiferromagnet}. Placing three such dots in a triangle is enough to realise a frustrated system, since it realises the prototypical example of frustration \index{frustration}: three antiferromagnetically \index{antiferromagnet} coupled (macro)spins. Such small clusters of dots were studied by the Paul Scherrer Institute group \cite{Mengotti2009}, who studied their energy spectra both experimentally and theoretically. Meanwhile, the Penn State group have studied (frustrated) kagome and (unfrustrated) honeycomb \index{honeycomb} lattices of such perpendicularly magnetised dots (depicted in panels (b) and (c) of Fig. \ref{geometries}) \cite{Zhang2012}, and found that there is a remarkable correspondence between the correlations in this type of kagome ice \index{kagome ice} and a conventional in-plane one, in spite of their very different magnetostatic \index{magnetostatic} couplings. This concept of building artificial systems from perpendicularly magnetised materials deserves more attention than it has received to date.

All of the lattices discussed so far have a single co-ordination number \index{co-ordination number}. For instance, in the square ice \index{square ice}, four islands meet at every vertex, whereas in the kagome ice \index{kagome ice}, the number of islands that interact at every vertex is three. In order to estimate the entropy of such systems (as Pauling did for the Ice I$_\mathrm{h}$ phase of H$_2$O \cite{Pauling1935}, and can easily be done for the six-vertex model of Lieb \cite{Lieb1967}), it is necessary only to extend the vertex-level degeneracy to the global ground state \index{ground state}. Nevertheless, it is simple to envisage lattices that have mixed co-ordination, where this basic procedure cannot be followed. This leads to a new class of vertex models where the frustration \index{frustration} arises not from the properties of an individual vertex but from the relative arrangements of neighbouring vertices, not all of which can take up their lowest energy configuration at the same time. Breaking the pure co-ordination \index{co-ordination number} rules allows for the construction of a wide variety lattices (with names such as the pinwheel, tetris \index{tetris}, or Santa Fe lattices) with extensive degeneracy \cite{Morrison2013}: the key property of the pyrochlore \index{pyrochlore} spin ice that is lost in the square ice \index{square ice}, with its twofold ground state \index{ground state}. The most heavily studied of these is the so-called shakti \index{shakti} lattice, constructed by removing elements from the square ice \index{square ice} system in such a way that there is a mixed 3/4 co-ordination. This has been shown theoretically to possess a quasicritical ice phase with algebraic correlations \cite{Chern2013}, similar to the critical correlations of the Coulomb phase in the pyrochlore \index{pyrochlore} spin ices. Importantly, this is the only lattice showing this new type of topologically emergent frustration \index{frustration} to have been experimentally realised \cite{Gilbert2014}. The results can be mapped onto the six-vertex model by breaking the lattice up into emergent composite plaquettes, which reveals the extensive degeneracy that it possesses. Whilst the four-island vertices possess no net magnetic charge in their ground state \index{ground state}, the three-island vertices necessarily must. These couple through long-range interactions that leads to crystallisation and screening. This idea of mixed co-ordination number \index{co-ordination number} lattices can also be used to construct not only emergent magnetic monopoles \index{monopole}, but also the associated emergent magnetic polarons \cite{Chern2013arxiv}. Screened charges have been directly visualised sa a transient state in dice lattice artificial spin ices \index{artificial spin ice} \cite{Farhannatcomm2016}. All of mixed co-ordination numner \index{co-ordination number} models, and the topologically emergent frustration \index{frustration} that they possess, can only be realised through the artificial spin ice \index{artificial spin ice} route, since they have no known analogs in nature, and remain a fruitful avenue for future work.

Spin ices form the archetype for geometrically frustrated magnetic systems, where the frustration \index{frustration} persists even in the limit of structurally perfect crystalline order. On the other hand, positional disorder/random interaction lead to another famous form of frustrated magnetic system: the spin glass \cite{MydoshBook}. Such systems naturally rely on a lack of crystalline order to provide the necessary random couplings. (The idea of an artificial analog of a spin glass is not one that seems to have been pursued yet.) At the boundary of crystals, with their perfect discrete translation symmetry, and glasses, with the total absence of any such symmetry, lie the quasicrystals \index{quasicrystal}. These possess order (in the sense that knowing a small part of the structure, one can follow rules to construct the rest of it and fill all of space) but lack translational symmetry. Quasicrystals \index{quasicrystal} containing both rare earth \cite{Islam1998} and transition metal \cite{Chernikov1993} magnetic species show spin glass-like freezing when the spins are dilute. Nevertheless, geometry appears to play some role. The discovery of quasicrystals \index{quasicrystal} \cite{Schectman1984} forced a redefinition by the International Union of Crystallography of the term ``crystal'' in 1991 to refer to structures that produce sharp diffraction patterns \index{diffraction pattern}, regardless of whether they posses discrete translational symmetry or not. Mathematically, it is now understood that quasicrystals \index{quasicrystal} are projections in three-dimensional space of structures that are periodic in some higher number of dimensions. This applies to magnetism as well as structure: the neutron diffraction pattern \index{diffraction pattern} of a Zn-Mg-Ho icosahedral quasicrystal \index{quasicrystal} can be understood in terms the system realising a projection of a higher dimensional multi-$q$ antiferromagnetic \index{antiferromagnet} structure \cite{Sato2000}.

The two-dimensional analog of a quasicrystal \index{quasicrystal} is a Penrose tiling \cite{Penrose1974}. Heisenberg spins on the nodes of a Penrose tiling have been studied theoretically \cite{Vedmedenko2003} and their configuration can be shown to correspond to interpenetrating non-collinear sublattices in a higher dimensional structure \cite{Vedmedenko2006}. It is natural to use these results at the inspiration to build an artificial analog of a magnetic quasicrystal \index{quasicrystal} using the same methodology as building an artificial spin ice \index{artificial spin ice}: place magnetic nanoislands with Ising-like \index{Ising} spins along the edges of a Penrose tiling and study their configurations.

Penrose patterns come in two forms, known as kite-and-dart or rhomboid, with the names derived from the shapes of the tiles used to form them. Bhat \textit{et al.} have constructed Permalloy \index{Permalloy} lattices based on the kite-and-dart form, where the elements are connected to form a continuous network \cite{Bhat2013}. These were studied using the macroscopic probes of SQUID magnetometry and FMR, revealing well-defined switching fields and a rich mode structure. However, magnetic imaging of the patterns was not reported. Meanwhile, Fig. \ref{geometries} shows a small portion of an artificial magnetic quasicrystal \index{quasicrystal} constructed in the author's laboratory built using discrete islands along the edges of rhomboid tiles, which take the forms of thin or fat rhombi. In this pattern, in contrast to the kite-and-dart, all the links between vertices of the tilings are the same length. Theoretical considerations and MFM \index{MFM} imaging show that the pattern contains rigid parts with a two-fold ground state \index{ground state} and loose spins that lead to extensive degeneracy, similar to the decagonal ordering in the Heisenberg system \cite{Vedmedenko2003}. The very high level of co-ordination number \index{co-ordination number} mixing--there are vertices where as few as three or as many as seven elements meet--leads to strong topologically induced frustration \index{frustration}. The enhanced level of frustration \index{frustration} means that the ground state \index{ground state} is very difficult to access, a phenomenon seen in a weaker form in the square ice \index{square ice} \cite{Li2010b}.

\section{From Effective to Real Thermodynamics}
\label{sec:thermo}

It has become a commonplace that artificial spin ices \index{artificial spin ice} are realisations of the toy models of statistical mechanics, such as the square ice \index{square ice} model \cite{Lieb1967}. What is missing so far from the discussion is one of the most important parameters of that sub-field of physics: temperature. So far in this chapter we have implicitly assumed that the state of the array that is being inspected through microscopy is fixed, so that we can acquire a stable image, and that any changes to that state are driven by the application of a field. What that means in practice is that $E_\mathrm{b} \gg k_\mathrm{B}T$ must hold, where $E_\mathrm{b}$ is the energy barrier between the bistable states of an island and $k_\mathrm{B}T$ is the thermal energy at the ambient temperature $T$. Indeed, for islands of the size studied by Wang \textit{et al.} \cite{Wang2006} can be shown to have $E_\mathrm{b} \sim 10^5$~K, far in excess of any realistically achievable experimental temperature. Thus thermal fluctuations \index{fluctuation} play no role.

Such a system is said to be athermal \index{athermal}. It is not ergodic \index{ergodic} and its states may not correspond to thermally equilibrated states. Such athermal \index{athermal} systems may nevertheless be described in terms of an effective thermodynamics, originally developed in the field of granular matter \cite{Jaeger1992}. In that field, the process of vibro-fluidisation is used to relax the many-body system into a low energy state \cite{2003DAnna}. The equivalent process in magnetism is an ac demagnetisation \index{demagnetisation}.

\begin{figure}[t]
  \includegraphics[width=8cm]{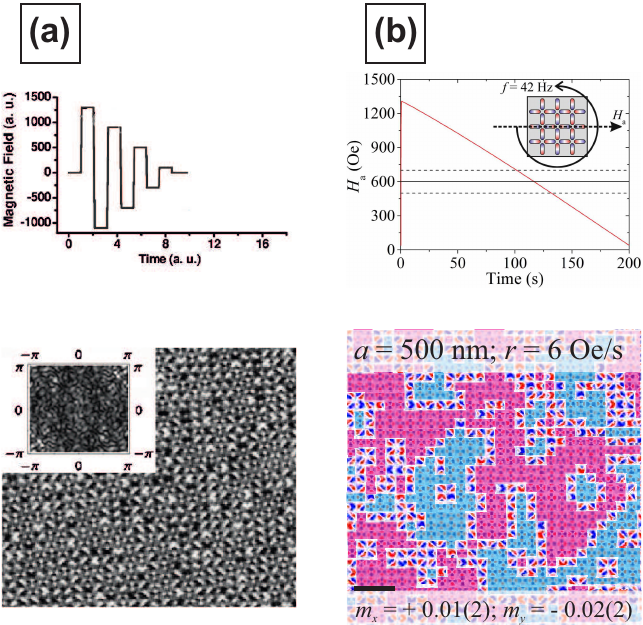}
  \caption{Rotating field demagnetisation protocols: (a) ac demagnetisation field profile (with results) (after Wang \textit{et al.} \cite{2007Wang}), and (b) linear demagnetisation field profile (with results) (after Morgan \textit{et al.} \cite{Morgan2013}).  \label{demag}}
\end{figure}

The demagnetisation \index{demagnetisation} protocol used in the initial experiments by Wang \textit{et al.} \cite{Wang2006} to relax the spin ices into a state where fidelity to the ice rules \index{ice rule} could be tested was to subject the sample to an ac demagnetisation \index{demagnetisation} process whilst rapidly spinning it at about 20~Hz so that the field samples all in plane directions in the array, as shown in Fig. \ref{demag}(a) and (b). The aim of a demagnetisation \index{demagnetisation} process is, or course, to reduce the magnetisation as close to zero as possible. There are of course, many such square ice \index{square ice} states with zero magnetisation: one is the antiferromagnetic \index{antiferromagnet} \index{ground state} formed from an alternating tiling of the two Type 1 vertices, but also many higher energy states with disordered moments. At first these protocols were able to provide this disorder \cite{2007Wang}, and thus degeneracy \cite{Nisoli2007}, but ground state \index{ground state} access was not possible. Given the energy scales involved, a state very close to the ground state \index{ground state} is expected to be the thermally equilibrated one. A more detailed study of the states yielded by this protocol showed that whilst it can be improved by using finer and finer field steps, the energy of the system does not extrapolate to the ground state \index{ground state} \cite{2008Ke}. Some years later, a linear demagnetisation \index{demagnetisation} protocol was shown to produce partial ground state \index{ground state} ordering (up to $\sim 60$ per cent of the vertices in the Type 1 configuration, as shown in Fig. \ref{demag}(c)) \cite{Morgan2013}, as previously predicted by theory \cite{2010Budrikis}. Slow field sweep rates are required, so that each island has a chance to experience all field directions in the few moments before it approaches its switching field. The degree of ground state \index{ground state} fidelity is constrained by the level of quenched disorder \cite{2012BudrikisMorgan}.

Nevertheless, these field driven protocols seem to be incapable of producing properly thermalised states, perhaps because they only permit downward steps in energy, rather than the upward and downward steps permitted by a true anneal \index{anneal} process. Nevertheless, the states they do produce can be described by an effective thermodynamics \cite{Nisoli2007}. Notions of the effective temperature \index{effective temperature} of the macrospin \index{macrospin} system, as introduced in the field of granular matter \cite{Jaeger1992}, can be brought into play \cite{Nisoli2010}. This effective temperature \index{effective temperature} is derived from an inspection of the artificial spin ice \index{artificial spin ice} of the system, and is thus a microscopic measurement, rather than the macroscopic measurements required to establish the real temperature. It is based on a simple vertex gas model in the canonical ensemble, where the energy of a vertex of Type $i$ is denoted by $E_i$, and its multiplicity (the number of equivalent configurations) by $\nu_i$. It is convenient to define an energy scale such that $E_1 = 0$ and $E_3 = 1$. It is worth noting that this formalism neglects any kind of correlations between adjacent vertices, even the most basic one that each macrospin \index{macrospin} must be a dipole. It is an effective thermodynamics since $\nu_2$, which ought to be equal to 4 (see Fig. \ref{square_ice}), is set to be 5. The reason for this is that an additional Type 2 vertex type is defined, which is the background configuration defined by the rotating field. In this state, all the macrospins \index{macrospin} track the field, rotating between the four different Type 2 configurations. As the field is reduced through the switching field distribution of all of the macrospins \index{macrospin}s in the array (which has a finite width due to both quenched disorder and the couplings between the islands) it carves defects into this background as each vertex drops out of the rotating background population into one of the four conventional vertex types. To determine the effective temperature \index{effective temperature} $T_\mathrm{eff} = 1/\beta_\mathrm{eff}$, one measures the fractional population $n_i$ of each vertex type (with $\sum_i n_i = 1$) and notes that this should be defined by a Boltzmann factor as $n_i = q_i \exp (-\beta_{\rm eff}E_i)/Z$, where $Z$ is the partition function. The ratio of any two of these $N_i$ will suffice to determine the effective temperature \index{effective temperature}, but the definition of zero energy means that it is convenient to compare any other vertex population to $n_1$. This leads to $\beta_\mathrm{eff}  = E_i^{-1} \ln \left[(q_i/2) (n_1/n_i) \right]$. Nisoli \textit{et al.} used this approach to show that when rotational ac demagnetisation \index{demagnetisation} is used, more strongly coupled arrays generally had lower effective temperatures \index{effective temperature} than more weakly coupled ones \cite{Nisoli2010}. In a truly thermally equilibrated system, the real and effective temperature \index{effective temperature}s will be the same, and any variation between them is a measure of how athermal \index{athermal} the systems is. Extending the idea of there being an effective statistical mechanics at play, it has also been shown to be possible to determine an effective entropy of the macrospin \index{macrospin} system from an inspection of the artificial spin ice \index{artificial spin ice}, once a suitable formalism has been developed \cite{2010Lammert}.

\begin{figure}[t]
  \includegraphics[width=10cm]{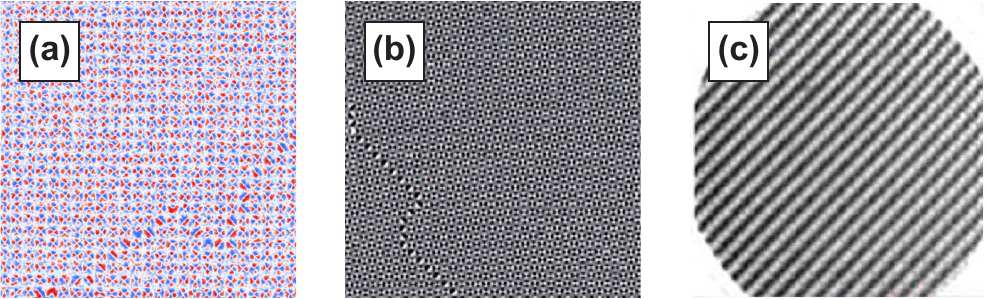}
  \caption{Thermally equilibrated artificial spin ices. (a) One shot thermalised, imaged using MFM. (b) Thermally annealed, imaged using MFM. After Zhang \textit{et al.} \cite{Zhang2013}. (c) Thermally relaxed, imaged using PEEM. After Farhan \textit{et al.} \cite{Farhan2013b}.\label{thermalisation}}
\end{figure}

Nevertheless, a true thermalisation \index{thermalisation} remained highly sought after, since (i) it would allow for the `real' thermodynamics of the system to be studied and (ii) obtaining a state close to the ground state \index{ground state} means that excitations \index{excitation} above it may be easily studied. A one shot thermalisation \index{thermalisation} process was discovered by Morgan \textit{et al.}, who found extensive ground state \index{ground state} order in some of their samples in their as-grown state after fabrication \cite{Morgan2011a}. An important point is that these samples were fabricated by liftoff, and so comprised discrete islands that passed through every thickness between zero and their final thickness during fabrication. The barrier to reversal $E_\mathrm{b}=KV$, where $K$ is an anisotropy constant defined by the island shape and $V$ is the island volume. When very thin, the island volume would be small and so $KV$ will be smaller than the thermal energy $k_\mathrm{B}T$ allowing free thermal flluctuation \index{fluctuation}. As the thickness increases, $KV$ will rise until it exceeds the thermal energy by a large enough factor that the system arrests, dropping out of thermal equilibrium and becoming athermal \index{athermal}. It thus captures a snapshot of a truly thermal state at the temperature and island thickness when this arrest takes place: an example is shown in Fig. \ref{thermalisation}(a). Theoretical treatments of this process were given by Nisoli \cite{Nisoli2012} and Greaves and Muraoka \cite{2012Greaves}. States with large domains of ground state \index{ground state} order, separated by domain walls \cite{Budrikis2012a}, and containing isolated excitations \index{excitation} were observed. These excitations \index{excitation} were distributed according to Boltzmann factors \cite{Morgan2011a}. These states had very low effective temperatures \index{effective temperature}, which could be described within a canonical ensemble model where $\nu_2$ could be set to the true value of 4 \cite{2013Morgan}. Thus, the effective temperature \index{effective temperature} model shows that truly thermalised states can be discriminated from those prepared by field demagnetisation \index{demagnetisation} on the basis of their statistical properties, just as anticipated by Ke \textit{et al.} \cite{2008Ke}.  Measurements of such states allow access to phase transitions between the antiferromagnetically \index{antiferromagnet} ordered ground state \index{ground state} and a high effective temperature \index{effective temperature} disordered state \cite{Levis2013}.

A drawback of this one-shot anneal \index{anneal} process is that it can only be performed during the sample growth, and the thermalised state is irrecoverably destroyed once the sample state is changed by a field, and the ground state \index{ground state} can never be reset. An obvious approach to this problem is to carry out a true anneal \index{anneal} of the sample, heating it to above the relevant temperature scale, which is either the blocking temperature $T_\mathrm{b}$ of the macrospins \index{macrospin} or the Curie temperature $T_\mathrm{c}$ of the microspins (whichever is lower), and then cooling back to room temperature in order to image the state that had been prepared. This was attempted by several groups, but in every case the nanostructures did not survive the heat treatment and the artificial spin ice \index{artificial spin ice} arrays were destroyed.

The first success in achieving anything resembling a reversible anneal \index{anneal} was made by Kapaklis \textit{et al.}, who fabricated artificial spin ice \index{artificial spin ice} arrays from atomic Fe monolayers sandwiched between Pd layers \cite{Kapaklis2012}. These had a Curie point below room temperature and so could be heated and cooled through $T_\mathrm{c}$ to `melt' and `freeze' the macrospin \index{macrospin} system. This was detected through MOKE magnetometry, but the states obtained were not imaged. An experiment along similar lines was carried out by Porro \textit{et al.}, who reduced the $T_\mathrm{c}$ of Permalloy \index{Permalloy} by enriching it with Ni \cite{Porro2012}. Although the reduction was fairly modest, it was sufficient to reduce the temperature needed for a magnetic anneal \index{anneal} to a low enough value that the artificial \index{artificial spin ice} islands survived. Extensive ground state \index{ground state} order was achieved in square ice \index{square ice} samples. The process was perfected by Zhang \textit{et al.}, who fabricated their artificial spin ices \index{artificial spin ice} on substrates coated with an inert layer of Si$_3$N$_4$. This allows for annealing \index{anneal} up to high enough temperatures to completely anneal \index{anneal} the macrospin \index{macrospin} system \cite{Zhang2013}. They studied both square and kagome ice \index{kagome ice} arrays, and were able to obtain perfect ground state \index{ground state} order in the former (see Fig. \ref{thermalisation}(b)), and to observe the charge crystallisation due to the interactions of the magnetic charges intrinsic to the odd coordination at each vertex predicted by theory \cite{2009Moller,Chern2011} in the latter. The same effect has achieved by Drisko \textit{et al.}, who used the low Curie temperature ferrimagnet FePd$_3$ ($T_\mathrm{c} \sim 200~^\circ$C to achieve perfect ground state \index{ground state} ordering in square ices \index{square ice} and an even higher degree of monopole \index{monopole} charge order in connected kagome systems \cite{Drisko2015}. This approach has been extended to the study of deliberately introduced topological defects--dislocations, which possess a Burgers vector--that disrupt the ability to form ground state \index{ground state} order by forcing into being an excited string of Type 2 and 3 vertices that stretches to the edge of the array \cite{Drisko2015arxiv}.

Once static thermalised states had been observed, the next step is clearly to observe thermally activated dynamics. The first hints of this were seen by Arnalds \textit{et al.}, where a handful of unstable islands were seen in otherwise frozen thermalised arrays \cite{2012Arnalds}. Fully fluctuating kagome clusters were soon achieved by making very thin--and hence low volume--Permalloy \index{Permalloy} islands \cite{Farhan2013a}, which can nevertheless be easily imaged by PEEM \index{PEEM} due to the surface sensitivity of that technique. Applying the same idea to full square ice \index{square ice} arrays led to a remarkable discovery: after saturation by a field into a fully Type 2 configuration the system is able to relax thermally into the ground state \index{ground state} by nucleating monopole-antimonopole \index{monopole} pairs, which create Dirac strings \index{Dirac string} that flip the correct chains of islands to yield Type 1 vertices \cite{Farhan2013b}, as shown in Fig. \ref{thermalisation}(c). That the interactions between islands help to stabilise their macrospins \index{macrospin} against thermal fluctuations \index{fluctuation} was shown by studying the dynamics in Fe-monolayer in Pd kagome ices \index{kagome ice} with different lattice spacings \cite{Kapaklis2014}. A beautiful recent result from Gilbert \textit{et al.} is the demonstration of emergent reduced dimensionality \index{dimensionality} in the tetris \index{tetris} lattice: on cooling from the fluctuating state, the system subdivides into oblique stripes of stable ground state \index{ground state} order interspersed with stripes of fluctuating spins \cite{Gilbert2015}.

\section{Outlook}
\label{sec:outlook}

The field of artificial spin ices \index{artificial spin ice} remains active with many new breakthroughs emerging. These include the cross-fertilisation with magnonics \index{magnonics} \cite{Gliga2013,Zhou2016,Bhat2016}, work on disorder-induced criticality \cite{Chern2014b}, proposals for three-dimensional systems \cite{Chern2014a}, and the development of complex network theories to represent the changing states in artificial spin ice \index{artificial spin ice} systems \cite{Budrikis2012b}.

Whilst nanomagnetic systems remain by far the most popular, the general concept of constructing artificial frustrated systems using nanotechnological means has found expressions in other forms. Superconducting \index{superconducting} systems have been studied in various forms, with the earliest implementations using coupled superconducting \index{superconducting} rings \cite{1996Davidovic,1997Davidovic}. A theoretical proposal for a square ice \index{square ice} in which each element contains a single superconducting \index{superconducting} vortex forms a very close analog with the frustrated proton system Ice I$_\mathrm{h}$, where the vortex must occupy one of two positions that are close to one vertex or the other, leading to a two-in/two-out ice rule \index{ice rule} at each vertex \cite{2009Libal}. Just a such a system was very recently realised (albeit using slightly different means to provide the bistable energy landscape for the vortices) by Latimer \textit{et al.} \cite{Latimer2013}. Soft matter approaches are also possible. A square ice \index{square ice} system where interacting colloidal \index{colloid} particles are held in optical traps was theoretically analysed by Lib\'{a}l \textit{et al.} \cite{2006Libal}, which operates on much the same principle as the superconducting \index{superconducting} vortex systems: the particle must lie at one end or the other of an elongated trapping potential, and thus be present at one or other vertex. Again, the analogy with protons in Ice I$_\mathrm{h}$ is very close. On the other hand, a realisation of the frustrated triangular antiferromagnet \index{antiferromagnet} was constructed by placing colloidal \index{colloid} particles between in a layer confined between two sheets such that the particles can move up or down a small distance \cite{Han2008}. By packing them a suitable density, complex patterns can be formed due to the frustration \index{frustration}. An advantage of this approach is that the colloidal \index{colloid} particles are large enough to image using an ordinary optical microscope. Going further in this direction, it is not even always necessary to use nanoscale systems: a macroscopic kagome spin ice was constructed by Mellado \textit{et al.} using centimeter scale bar magnets on hinged mounts, the dynamics of which were studied using a high speed video camera \cite{Mellado}.

From a more technological perspective, viewing these systems as information storage \index{information storage} \cite{Lammert2010,Wang2016} or processing technologies could also lead to nanomagnetic logic \index{logic} architectures \cite{Cowburn2000,Imre2006} based on frustrated arrays. These need not have the traditional von Neumann architecture: for instance, the Ising \index{Ising} model provides the theoretical underpinning for the Hopfield model of neural networks \index{neural network} \cite{Hopfield1982}, suggesting that non-Boolean neuromorphic \index{neuromorphic} architectures based on artificial spin ices \index{artificial spin ice} might be possible. Advances in the field of nanomagnetic logic also cross-pollinate into new experimental opportunities in artificial spin ices \index{artificial spin ice}, for instance driving out-of-equilibrium dynamics using spin Hall torques \cite{Bhowmik2014}. Whilst initially perceived as an attempt simply to build analogs of existing systems such as pyrochlores \index{pyrochlore}, new physics, such as topologically emergent frustration \index{frustration}, can only be studied by this route, which is an important justification for continued work in the field. It is clear that there will be plenty more opportunities to put the drinks on ice in order to celebrate future breakthroughs.

\section{Acknowledgements}
\label{sec:acknowledgements}

I am grateful to Jason Morgan for the AFM/MFM images in Fig. \ref{square_ice} and MFM images in Fig. \ref{micrographs}(a) and Fig. \ref{thermalisation}(a), Aaron Stein for the SEM image in Fig. \ref{geometries}(f), and Sophie Morley for the AFM/MFM images in Fig. \ref{kagome_ice}, and (with assistance from Peter Fischer and Mi-Young Im) for the XTM image in Fig. \ref{micrographs}(d).

\bibliographystyle{spphys}

\bibliography{ASI}

\printindex

\end{document}